\documentclass[11pt]{article}
\usepackage{amsmath,amssymb}

\makeatletter
\@addtoreset{equation}{section}
\makeatother

\def\ben{\begin{equation}}
\def\een{\end{equation}}

  \let\n=\nu  \let\p=\pi

\let\C=\Chi

\def\nn{\nonumber} \def\bd{\begin{document}} \def\ed{\end{document}}
\def\ds{\documentstyle} \let\fr=\frac \let\bl=\bigl \let\br=\bigr
\let\Br=\Bigr \let\Bl=\Bigl
\let\bm=\bibitem
\let\na=\nabla
\let\pa=\partial \let\ov=\overline
\newcommand{\be}{\begin{equation}}
\newcommand{\ee}{\end{equation}}
\def\ba{\begin{array}}
\def\ea{\end{array}}
\def\ft#1#2{{\textstyle{\frac{\scriptstyle #1}{\scriptstyle #2} } }}
\def\fft#1#2{{\frac{#1}{#2}}}
\def\del{\partial}
\def\vp{\varphi}
\def\sst#1{{\scriptscriptstyle #1}}
\def\oneone{\rlap 1\mkern4mu{\rm l}}
\def\td{\tilde}
\def\wtd{\widetilde}
\def\ie{{\it i.e.\ }}
\def\dalemb#1#2{{\vbox{\hrule height .#2pt
        \hbox{\vrule width.#2pt height#1pt \kern#1pt
                \vrule width.#2pt}
        \hrule height.#2pt}}}
\def\square{\mathord{\dalemb{6.8}{7}\hbox{\hskip1pt}}}
\newcommand{\ho}[1]{$\, ^{#1}$}
\newcommand{\hoch}[1]{$\, ^{#1}$}
\newcommand{\bea}{\begin{eqnarray}}
\newcommand{\eea}{\end{eqnarray}}
\newcommand{\ra}{\rightarrow}
\newcommand{\lra}{\longrightarrow}
\newcommand{\Lra}{\Leftrightarrow}
\newcommand{\bp}{\tilde \beta^\prime}
\newcommand{\tr}{{\rm tr} }
\newcommand{\Tr}{{\rm Tr} }
\def\0{{\sst{(0)}}}
\def\1{{\sst{(1)}}}
\def\2{{\sst{(2)}}}
\def\3{{\sst{(3)}}}
\def\4{{\sst{(4)}}}
\def\5{{\sst{(5)}}}
\def\6{{\sst{(6)}}}
\def\7{{\sst{(7)}}}
\def\8{{\sst{(8)}}}
\def\n{{\sst{(n)}}}
\def\cA{{{\cal A}}}
\def\cB{{{\cal B}}}
\def\cF{{{\cal F}}}
\def\cG{{{\cal G}}}
\def\cH{{{\cal H}}}
\def\tV{\widetilde V}
\def\tW{\widetilde W}
\def\tH{\widetilde H}
\def\tE{\widetilde E}
\def\tF{\widetilde F}
\def\tA{\widetilde A}
\def\im{{{\rm i}}}
\def\tY{{{\wtd Y}}}
\def\ep{{\epsilon}}
\def\vep{{\varepsilon}}
\def\bD{{{\bar D}}}
\def\R{{{\mathbb R}}}
\def\C{{{\mathbb C}}}
\def\H{{{\mathbb H}}}
\def\CP{{{\mathbb C}{\mathbb P}}}
\def\RP{{{\mathbb R}{\mathbb P}}}
\def\Z{{{\mathbb Z}}}
\def\bA{{{\mathbb A}}}
\def\bB{{{\mathbb B}}}
\def\bC{{{\mathbb C}}}
\def\bD{{{\mathbb D}}}
\def\bE{{{\mathbb E}}}
\def\bZ{{{\mathbb Z}}}
\def\Re{{{\frak{Re}}}}
\def\Im{{{\frak{Im}}}}
\def\cosec{{\,\hbox{cosec}\,}}
\def\Gm{{\Gamma_{\!\! -}}}
\def\Gp{{\Gamma_{\!\! +}}}
\def\stan{{standard }}
\def\nonstan{{supernumerary }}
\def\p{{\partial}}
\def\kdel#1{{\fft{\del}{\del#1}}}

\def\bog{{Bogomolny }}
\def\om{{\omega}}

\newcommand{\tamphys}{\it George and Cynthia Woods Mitchell  Institute
for Fundamental Physics and Astronomy,\\
Texas A\&M University, College Station, TX 77843, USA}

\newcommand{\auth}{
H. L\"u\hoch{\dagger,\ddagger} and  Jianwei Mei\hoch{\dagger}
}

\begin{document}

\begin{flushright}
\hfill{
MIFP-08-15\ \ \ USTC-ICTS-08-11}\\
\end{flushright}

\begin{center}

{\large {\bf Ricci-Flat and Charged Wormholes in Five Dimensions
}}

\vspace{25pt}

\auth

\vspace{10pt}
\hoch{\dagger}{\tamphys}

\vspace{10pt}

\hoch{\ddagger}{\it Interdisciplinary Center of Theoretical Studies,
USTC, Hefei, Anhui 230026, PRC}

\vspace{25pt}

\underline{ABSTRACT}

\end{center}

\vspace{15pt}

We construct stationary Ricci-flat inter-universe Lorentzian wormhole
solutions in all $D\ge 5$ dimensions that connect two flat asymptotic
spacetimes.  Such a solution can be viewed as the gravity dual of a
string tachyon state whose linear momentum is larger than its
tension. We focus our analysis on the $D=5$ wormholes which are not
traversable for the timelike and null geodesics; however, we
demonstrate that there exist accelerated timelike trajectories that
traverse from one asymptotic region to the other.  We further study
the minimally-coupled scalar wave equation and demonstrate that the
quantum tunnelling between two worlds must occur.  We also obtain
charged wormholes in five-dimensional supergravities. With appropriate
choice of parameters, these wormholes connect AdS$_3\times S^2$ in
one asymptotic region to flat Minkowskian spacetime in the other.

\thispagestyle{empty}

\pagebreak
%
%


\section{Introduction}

Einstein's discovery of general relativity has since inspired people's
imagination of taking a ``shortcut'' in spacetime travel.  The first
non-trivial wormhole construction is called the Einstein-Rosen bridge
that combines a Schwarzschild black hole and a white hole \cite{er},
which unfortunately turns out to be rather short lived \cite{fw}.
However, the existence of wormhole is not inconsistent with general
relativity.  Most of the studies of this subject have been focused on
four dimensions, with either Lorentzian or Euclidean signatures.
Typically a wormhole requires a rather unusual energy-momentum tensor
to maintain.

With the advent of string theory that demands higher spacetime
dimensions, it is natural to examine the possibility of wormholes
beyond four dimensions.  Euclidean wormholes in string theory have
recently been studied \cite{mm,bcpvv,hop,bd,bcptv} as solutions of
supergravities.  In this paper, we construct stationary Ricci-flat
inter-universe Lorentzian wormholes in all $D\ge 5$ dimensions.
The solution is presented in section 2. Using $D=5$ as an example,
we demonstrate that it is a smooth solution that links two
asymptotic flat Minkowskian spacetimes. We find that the solution
can be viewed as a gravity dual of a string tachyon state that has
a linear momentum larger than its tension. The solution has a BPS
limit that leads to a supersymmetric pp-wave, and the two worlds
then become disconnected by the pp-wave singularity.

We analyze the geodesic and non-geodesic motions in section 3. The
wormholes are not traversable for the timelike or null geodesics.
However, we obtain accelerated timelike trajectories that can
traverse from one asymptotic region to the other.  We demonstrate
that the maximum acceleration along the trajectories can be
arbitrarily small, and the total impulse required to cross over
is finite, and independent on the size of the wormhole.

We then consider the quantum tunnelling of the wormholes in section 4
by examining the minimally-coupled scalar wave equation.  We
demonstrate that the radial wave equation can be reduced to a
one-dimensional quantum mechanical system with a finite potential
barrier.  This implies that the quantum tunnelling must take place.  We
obtain the approximate transmission rate using the delta function
approximation for the potential.  The transmission rate is related to
the size of the wormhole, and it vanishes when the wormhole vanishes.

In section 5, we obtain both electrically and magnetically charged
wormhole solutions in five-dimensional supergravities.  The later one
can be viewed as the magnetically charged tachyonic string.  Although
the solutions become more complicated, the essential feature of the
neutral Ricci-flat wormhole remains.  With appropriate choice of
parameters, we find the charged wormholes that connect AdS$_3\times
S^2$ in one end to flat Minkowskian spacetime in the other.

We conclude the paper in section 6. And finally in the appendix, we
present the general wormhole solution in five-dimensional $U(1)^3$
supergravity that carries both electric and magnetic charges.

\section{The metric}

The Ricci-flat Lorentzian wormhole solution in all dimensions
$D\geq 5$ is given by
\be ds^2= (r^2 + a^2) d\Omega_{D-3}^2 + \fft{r^2
dr^2}{(r^2+a^2)\sin^2 h(r)} -\cos(\gamma + f) dt^2 +\cos(\gamma-
f)dz^2+ 2\sin f\, dt dz\,,\label{generalD} \ee
where
\be h(r)=\arctan\sqrt{\left(1+\fft{r^2}{a^2}\right)^{D-4} -1}\
,\qquad f=\fft{2\pi - \sqrt{2D-6}\, h(r)}{\sqrt{D-4}}\,. \ee
Here $d\Omega_{D-3}^2$ is a metric for the unit $S^{D-3}$, and
$\gamma$ and $a$ are constants.  It is clear that the case of $D=5$ is
particularly simple, and it was previously obtained in \cite{cd} and
further studied in \cite{aagc,aagccf}.  For simplicity, we shall focus
our attention on $D=5$ and examine the properties that are important
for our purpose.

We find that the metric for the Ricci-flat wormhole solution in five
dimensions can be written as follows
\be
ds^2=dr^2 + (r^2 + a^2) d\Omega_2^2 - \fft{r^2-2as\, r -
a^2}{r^2 + a^2} dt^2 -\fft{4ac\,r}{r^2+a^2} dt\,dz +
\fft{r^2+2a s\, r-a^2}{r^2+a^2} dz^2\,,\label{d5met1}
\ee
where $d\Omega_2^2=d\theta^2 + \sin^2\theta\, d\phi^2$ is the metric
for a unit $S^2$, and $c=\cosh\beta$, $s=\sinh\beta$ and $a$ are real
constants, which, without loss of generality, are taken to be
non-negative. Note we have made certain coordinate rotation and
reparametrisation to obtain (\ref{d5met1}) from (\ref{generalD}).  The
off-diagonal term in the metric implies that the solution is not
static, but stationary.

   There are two asymptotic regions, corresponding to $r\rightarrow
\pm \infty$, in which the metric has the following form
\be
ds^2=-dt^2 + dz^2 + dr^2 + r^2 d\Omega_2^2\,.
\ee
This describes (Minkowskian)$_5$ if $z$ is a real line, or
(Minkowskian)$_4\times S^1$ if $z$ is a circle.  We denote the
$r=\pm\infty$ worlds as W$_+$ and W$_-$.  There is no curvature
singularity in the bulk between the two worlds.  This can be seen from
the Riemann square, given by
\be
\hbox{Riem}^2 = \fft{24a^2 (a^2-2r^2)}{(r^2+a^2)^4}\,.
\ee
The two Riemann cubics can also be easily obtained, given
by\footnote{For the Schwarzschild-Tangherlini black hole, the
ratio $\hbox{Riem}^3/[\hbox{Riem}^2]^{3/2}$ is a pure numerical
constant.  Thus, our metric is not locally the black hole solution.
We are grateful to Chris Pope for this observation.}
\bea
\hbox{Riem}_1^3 &\equiv & R^{\mu}{}_{\nu}{}^\rho{}_\sigma\,
R^{\nu}{}_\mu{}^\alpha{}_{\beta}\, R^\beta{}_{\alpha}{}^\sigma{}_\rho=
\fft{192a^4 r^2}{(r^2+a^2)^6}\,,\nn\\
\hbox{Riem}_2^3 &\equiv & R^{\mu}{}_{\nu}{}^\rho{}_\sigma\,
R^{\alpha}{}_\mu{}^\beta{}_{\rho}\, R^\nu{}_{\alpha}{}^\sigma{}_\beta=
-\fft{96a^4 r^2}{(r^2+a^2)^6}\,,
\eea
The absence of the parameter $\beta$ in the Riemann curvature implies
that it is trivial.  Indeed, we can make the following coordinate
(un)boost,
\be
t\rightarrow \cosh\ft12\beta\, t + \sinh\ft12\beta\, z\,,\qquad
z\rightarrow \sinh\ft12\beta\, t + \cosh\ft12\beta\, z\,,
\ee
the metric becomes $\beta$ independent, given by
\be
ds^2=dr^2 + (r^2 + a^2)\,d\Omega_2^2 - \fft{r^2-a^2}{r^2+a^2} dt^2 -
\fft{4a\,r}{r^2 + a^2}dt\,dz + \fft{r^2 -a^2}{r^2+a^2} dz^2\,.
\ee
However, the off-diagonal term, which is crucial for the regularity of
the solution, can not be removed by any real coordinate
transformation.  We can instead make a complex coordinate
transformation
\bea%
    t &\longrightarrow&
\sqrt{\frac{1-\im\, s}{2}}~x+\sqrt{\frac{1+\im\,s}{2}}~z~~~,~~~r
\longrightarrow\im\,\tilde{t}~,\nonumber\\
    z &\longrightarrow&
\frac{s-\im}{c}\sqrt{\frac{1-\im\,s}{2}}~x+
\frac{s+\im}{c}\sqrt{\frac{1+\im\,s}{2}} ~z~,
\eea%
the metric (\ref{d5met1}) then becomes diagonal, namely
\be
ds^2 = -d\td t^2 + (a^2-\td t^2) d\Omega_2^2 +
\fft{a-\td t}{a+\td t} dx^2 + \fft{a+\td t}{a-\td t} dz^2\,.
\ee
This singular metric is a five-dimensional analogue of the Kasner
universe in $D=4$.

Although $\beta$ is a trivial boosting parameter, we shall keep it
in our discussion in this section, since it makes it easier to
take the BPS limit and to obtain the supersymmetric
purely-gravitational pp-wave.

    The metric (\ref{d5met1}) has no degenerate surface associated with
the radial coordinate $r$.  The radius of the $S^2$ never vanishes,
with the minimum being $a$, measuring the size of the wormhole.  The
determinant of the metric in $(t,z)$ directions is minus 1, {\it i.e.}
\be
\hbox{det}(g_{ij})=-1\,,\qquad\hbox{for}\qquad
i,j=\{t,z\}\,.
\ee
The metric has no globally defined time coordinate.  Whilst $t$ is a
good time coordinate for large absolute values of $r$, its r\^ole of
time is relegated to $z$ inside the wormhole.  There is an ergo region
\be
a(s-c)\le r\le a (s + c)\,,
\ee
in which $g_{tt}\ge 0$.  In the region
\be
-a(s+c)\le r\le a (c-s)\,,
\ee
we have $g_{zz}\le 0$.  For the choice of $z$ being circular, this
implies a closed-timelike-circle (CTC); for it being a real line, the
metric has no CTC.  One might think that a linear combination of
coordinates $t$ and $z$ could have a definite signature.  This is not
the case.  Consider the Killing vector $\ell=\mu\, \del/\del t+ \nu\,
\del/\del z$.  It is straightforward to obtain
\be
r=\pm \infty:\qquad
\ell^2 = - \mu^2 +\nu^2\,,\qquad\qquad
r=0:\qquad \ell^2=\mu^2-\nu^2\,.
\ee
It is perhaps instructive to use the light-cone coordinates
\be
t- z = u\,,\qquad t+z=v\,,
\ee
in which case, the metric becomes
\be
ds^2=\fft{a(c+s)r}{r^2 + a^2} du^2 -
\fft{r^2-a^2}{r^2+a^2} du\,dv -
\fft{a(c-s)r}{r^2 + a^2} dv^2 +
dr^2 + (r^2 + a^2) d\Omega_2^2\,.\label{lightcone}
\ee
The Killing vector $\del/\del u$ is spacelike in the region $r\in
(0,+\infty)$, timelike in $(-\infty,0)$ and null at $r=0$ and $r=\pm
\infty$.  The Killing vector $\del/\del v$, on the other hand, is
timelike in $(0,\infty)$, spacelike in $(-\infty,0)$ and also null at
$r=0$ and $r=\pm \infty$.

By studying the sub-leading structure of the metric at the infinities,
we obtain the mass and linear momentum per unit $z$ length, measured
in either the W$_+$ or the W$_-$ world.  They are given by
\be
M_\pm = \fft{3}{32\pi} \int_{r\rightarrow \pm \infty} *dK_t =
\pm \ft34 a\,s\,,\qquad
P_{\pm} = \fft{3}{32\pi} \int_{r\rightarrow \pm \infty} *dK_z=
\pm \ft34 a\,c\,.
\ee
Here $K_t$ and $K_z$ are 1-forms associated with the Killing vectors
$\del/\del t$ and $\del/\del z$ respectively.  Thus the wormhole is
created by an infinitely-stretched string (or a finite closed string)
with momentum along the string.\footnote{It is perhaps more
  appropriate to call the solutions wormstring or wormtubes; however,
  we shall continue to use wormholes, for the disinclination of
  inventing new words.}  The mass, or more precisely the tension of
the string, measured in the $r\rightarrow +\infty$ world W$_+$ is
positive, but it is negative in the $r\rightarrow -\infty$ world
W$_-$.  The evasion of the positive-energy theorem in the W$_-$ world
may be related to the fact that it is asymptotic
(Minkowskian)$_4\times S^1$.  Even if the coordinate $z$ is infinitely
stretched, it is singled out from the rest spatial coordinates near
the asymptotic infinities.  This is analogous to the Atiyah-Hitchin
metric whose Euclidean ``mass'' is negative.  The metric is
nevertheless smooth without violating the positive energy theorem due
to the fact that its asymptotic infinity is $\R^3\times S^1$, rather
than $\R^4$.  It may also be related to the fact that we now have two
asymptotic boundaries.  It is instructive to note that we have
$M_++M_-=0$, which is consistent with that our solution is
sourceless.

       Another property of the wormhole is that
\be
M^2 - P^2 = -\fft{9}{16} a^2 <0\,.
\ee
This implies that the string is of tachyonic nature.  Our wormhole
solution may be viewed as a gravitational dual of a string tachyon
state. In one limit, we can let $\beta=0$ and hence $s=0$ and $c=1$,
corresponding to a tensionless string with non-vanishing momentum.  In
the other limit, we can send $\beta\rightarrow \infty$ and
$a\rightarrow 0$, but keeping $a\,s\rightarrow q$ non-vanishing.  In
this limit, the BPS condition $M=P$ is satisfied.  The resulting
metric following from (\ref{lightcone}) describes a supersymmetric
pp-wave, given by
\be
ds^2=-du dv + \fft{2q}{r} du^2 + dr^2 + r^2 d\Omega_2^2\,.
\ee
The two worlds W$_+$ and W$_-$ are disconnected in this limit at
$r=0$, which is a spacetime singularity.

We can perform the Kaluza-Klein reduction on the string direction $z$,
and obtain an electrically-charged particle-like solution in four
dimensions.  The charge is larger than its mass, as in the case of all
the charged particles observed in our universe.  The solution is
singular.   It is rather common in string and supergravities that
lower-dimensional singular solutions become regular after they are
lifted up to higher dimensions \cite{ght}.

Although we have concentrate the analysis on the $D=5$ solutions, The
essential properties of the wormholes are shared by all higher
dimensional solutions in (\ref{generalD}). All these metrics for $D\ge
5$ describe stationary inter-universe Lorentzian wormholes. Each of
them have two asymptotic flat regions at $r=\pm \infty$ and the bulk
between these two regions has no singularity. There is no metric
singularity associated with the radial coordinate $r$.  The
determinant of the metric in $t$ and $z$ directions is minus one.
There is no globally defined time coordinate, and there exist BPS
limits for the wormholes to become supersymmetric pp-waves for all
dimensions.

    To conclude this section, we note that the metric (\ref{d5met1})
can also be put in the following natural vielbein base
\be ds^2=\fft{- [r (c dt - s dz) + a dz]^2 +
 [r (c dz - s dt) - a dt]^2}{r^2+a^2} +
dr^2 + (r^2 + a^2) d\Omega_2^2\,. \ee

\section{Geodesic motions and timelike trajectories}

The geodesic motions for the $D=5$ wormhole solution were discussed in
\cite{aagc}, and it was demonstrated that it is not geodesically
traversable.  In this section, we demonstrate that there exist
accelerated time trajectories that cross over the wormhole and
that the required total impulse is finite.

We begin by reviewing the geodesic motions, whose equations can be
obtained from the Hamilton-Jacobi equations; they can also be
equivalently derived from the Lagrangian
\bea
{\cal L} &=& \ft12 g_{\mu\nu} \dot x^\mu \dot x^\nu
= \ft12 \dot r^2 + \ft12 (r^2 + a^2) (\dot \theta^2 +
\sin^2\theta\, \dot\phi^2)\nn\\
&&\qquad\qquad - \fft{r^2- a^2}{2(r^2 +a^2)}\dot t^2-
\fft{2a\,r}{r^2+a^2} \dot t\,\dot z +
\fft{r^2 - a^2}{2(r^2 + a^2)} \dot z^2\,,
\eea
where a dot denotes a derivative with respect to the geodesic
parameter $\lambda$.  Note that we set the trivial parameter
$\beta=0$ in this section.

    The equations of motion for $\theta$ and $\phi$ admit the
following solution
\be
\theta=\ft12 \pi\,,\qquad \dot \phi=\fft{J}{r^2+a^2}\,,
\ee
where $J$ is an integration constant describing the orbiting angular
momentum of the geodesic particle, the test particle travelling along
the geodesics.  The equations of motion for $t$ and $z$ imply that
\be
\fft{r^2-a^2}{r^2 + a^2}\, \dot t +\fft{2a\,r}{r^2+a^2}
\,\dot z=E\,,\qquad
\fft{r^2-a^2}{r^2 + a^2}\, \dot z - \fft{2a\,r}{r^2+a^2}
\dot t=p\,,
\ee
where the integration constant $E$ and $p$ can be viewed as the energy
and the linear momentum in the $z$ direction of the geodesic particle.
The Lagrangian itself is a constant for geodesic motions, {\it i.e.},
${\cal L}=-\ft12\epsilon$, where $\epsilon=1,0,-1$ for the timelike,
null and spacelike geodesics respectively.   It follows that
we have
\be
\dot r^2 = -\epsilon - \fft{J^2}{r^2 +a^2} + V\,,
\ee
where the potential function $V$ is given by
\be
V=\fft{(E^2-p^2) (r^2-a^2) - 4 a\, E\,p\,r}{r^2 +a^2}
\,.
\ee
For the particle to exist at asymptotic regions, we must have that
$E^2-p^2 -\epsilon \ge 0$.  The potential $V$ has a minimum at
$r=a\,p/E$, with
\be
V_{\rm min} = -E^2 - p^2\,.
\ee
Thus the wormhole is not traversable for both the timelike and
null geodesics, corresponding to $\epsilon=1$ and 0 respectively.
For the spacelike geodesics, corresponding to $\epsilon=-1$, the
quantity $\dot r$ can stay real for all real values of $r$ provided
that $E^2+p^2\le 1$.  Although we present here the explicit analysis
for only the simplest case with $\beta=0$, the general feature, namely
that the wormholes are not traversable for the timelike and null
geodesics, remains also true for all $\beta\ne0$

Having established the geodesic non-traversability, we now consider
non-geodesic timelike trajectories.\footnote{We are grateful to Don
Page for part of this discussion.} For $\beta=0$, the metric
(\ref{lightcone}) can be written as
\be
ds^2=-\fft{(r\,du + a\,dv)(r\,dv - a\, du)}{r^2+a^2} +
dr^2 + (r^2 + a^2)\,d\Omega_2^2\,.
\ee
A timelike trajectory moving forward in time for large $r$ then requires,
for each positive $d\tau$ where $\tau$ is the proper time,
that $r\,du + a\,dv >0$ and $r\,dv - a\,du>0$.  Such a trajectory
does exist.  For example, let us consider
\be
u=\fft{\eta\,(r-a)^2}{a}\,,\qquad v=\fft{\eta\,(r+a)^2}{a}\,,
\ee
where $\eta$ is a positive constant.  It follows that
\be
r\, u' + a\, v' = r\, v' - a\, u'=\fft{2\eta\,(r^2+a^2)}{a}>0\,,\qquad
(r\,u' + a\,v')(r\,v'-a\,u') > (r^2+a^2)\,,
\ee
for all real $r$ provided that
\be
\eta>\fft12\,.
\ee
This leads to a timelike trajectory going from $r=+\infty$ to
$r=-\infty$.  To see this in more detail, let us consider
the case with $\theta=\ft12\pi$ and $\phi=0$.  The trajectory
motion is then governed by
\be
-\fft{(r\,\dot u + a\, \dot v) (r\,\dot v - a\, \dot u)}{r^2+a^2}
+\dot r^2 = -1\,.
\ee
Thus we have
\be
\dot r\, \sqrt{\fft{4\eta^2}{a^2}\, r^2 + (4\eta^2-1)} =-1\,.
\ee
It is clear that the velocity starts with zero at $r=+\infty$ and
reaches its maximum $1/\sqrt{4\eta^2-1}$ at $r=0$ and reduces to zero
again when $r=-\infty$.  The acceleration is given by
\be
\ddot r=-\fft{4\eta^2 a^2 r}{(4\eta^2 r^2 + a^2 (4\eta^2-1))^2}\,.
\ee
Having obtained the comoving velocity and acceleration, it is
straightforward to calculate the proper acceleration, given by
\be
A^\mu = \dot U^u + \Gamma^u{}_{\nu\rho} U^\nu\, U^\rho\,,
\ee
where $U^\mu=\dot x^\mu$ for the coordinates
$x^\mu=(u,v,r,\theta,\phi)$.  We find that
\be
A^2\equiv g_{\mu\nu}\, A^\mu A^\nu =
\fft{4\eta^2 a^2 \Big( (4\eta^2 +1)r^2 + a^2(4\eta^2-1)\Big)}{
(r^2+a^2)(4\eta^2 r^2 + a^2 (4\eta^2-1))^3}\,.
\ee
Thus the maximum acceleration occurs at $r=0$, given by
\be
A^2|_{\rm max} = \fft{4\eta^2}{a^2 (4\eta^2-1)^2}\,.
\ee
The constraint $\eta>\ft12$ implies that for the timelike trajectories
specified by the parameter $\eta$, the minimum value of maximum
acceleration can be arbitrarily small, since $\eta$ can be arbitrarily
large.

We now examine the total proper impulse that is required for
maintaining such an accelerated timelike trajectory.  This is given by
\bea
{\cal I}=\int A\, d\tau &=& \int_{\infty}^{-\infty}
\fft{A}{\dot r} dr\nn\\
&=&\int_{-\infty}^\infty \fft{2a\eta \sqrt{(4\eta^2+1)r^2 +
a^2(4\eta^2-1)}}{
(4\eta^2 r^2 + a^2(4\eta^2-1)) \sqrt{r^2+a^2}}\,dr\,.\label{energy}
\eea
This is finite because the integrand convenges as $1/r^2$ when
$r$ approaches $\pm\infty$.  The minimum impulse that is required for
these accelerated timelike trajectories is achieved when
$\eta\rightarrow \infty$.  In this case, we have
\be
{\cal I}_{\rm min}=
\int A\, d\tau = \int_{-\infty}^\infty \fft{a}{r^2 + a^2}\, dr=
\pi\,.
\ee
It is worth noting, following from (\ref{energy}), that the total
impulse required for crossing over the wormhole is independent on the
size of the wormhole.  This of course can be expected from the
dimensional analysis.

\section{Scalar wave propagation}

Since particles under the pure gravitional influence without
acceleration cannot pass the wormhole classically, we consider now the
quantum tunnelling effects by studying the minimally-coupled scalar
(of mass $m$) wave equation, namely
\be
\fft{1}{\sqrt{-g}} \del_{\mu}(\sqrt{-g} g^{\mu\nu} \del_{\nu} \Phi)
- m^2 \Phi=0\,.\label{scalar1}
\ee
We consider a radial wave of frequency $\omega$ and linear
momentum $p$, {\it i.e.},
\be
\Phi=\fft{e^{\im\,\omega\, t + \im\,p\,z} P_{\ell} (\cos\theta)}{\sqrt{r^2+a^2}}\,
u(r)
\ee
The scalar wave equation is then reduced to
\be
u'' - V_{\rm eff}\, u=0\,,
\ee
with the effective potential given by
\be
V_{\rm eff} = -(\omega^2-p^2-m^2) + \fft{a^2}{(r^2+a^2)^2} +
\fft{\ell(\ell+1) + 2a^2 (\omega^2-p^2)-4p\,\omega\,a\,r}{r^2 + a^2}
\,.
\ee
At asymptotic regions, the effective potential becomes a negative
constant $-(\omega^2-p^2-m^2)$. Defining $k=\sqrt{\omega^2-p^2-m^2}$,
the wave solutions at $r=\pm \infty$ have the following form
\be
u_\pm = A_\pm e^{\im\, kr} + B_\pm e^{-\im\, kr}\,.
\ee
Thus the wave equation can be reduced to a one-dimensional quantum
mechanical tunnelling system with a finite (non-singular) potential
barrier.  At $r=0$, the effective potential is given by
\be
V_{\rm eff}(r=0)=m^2 +\omega^2-p^2 + \fft{\ell(\ell+1)+1}{a^2} > 0\,.
\ee
This is consistent with the earlier result that the wormhole is
not traversable for geodesic motions.

     To illustrate that the tunnelling takes place and obtain the
transmission rate, let us consider a simpler case with $p=0$.  The
effective potential resembles a square potential with height
$(\ell(\ell+1) + 2 a^2\omega^2)/a^2$ and characteristic width $a$.  We
can use the delta function approximation, and write
\be
V_{\rm eff} \sim  - k^2 + S_V \delta(r),\quad
\hbox{with}\,,\quad
S_V \equiv\int^{+\infty}_{-\infty} (V_{\rm eff} + k^2)\, dr=
\fft{\pi}{2a} (1 + 2\ell(\ell+1) + 4a^2\omega^2)\,.
\ee
For such a potential, we may use the standard ansatz
\be
u(r<0)=c_T\, e^{\im k r}\,,\qquad
\hbox{and}\qquad u(r>0) = e^{\im k r} + c_R\, e^{-\im k r}
\ee
to obtain
\be
c_R=-\fft{1}{1 - \im k/S_V}\,,\qquad
c_T=-\fft{\im k/S_V}{1 - \im k/S_V}\,.
\ee
Hence the reflection and the transmission rates are given by
\be
R\sim 1 - \fft{k^2}{S_V^2}\,,\qquad
T\sim \fft{k^2}{S_V^2}=\fft{4(ka/\pi)^2}{(1+2\ell(\ell+1) +
4a^2\omega^2)^2}
\,.
\ee
Note that the delta function approximation is best justified when the
radial wave length is much greater than the characteristic width of the
potential barrier, {\it i.e.} $ka<<1$.  The result shows that the
transmission rate is smaller for the smaller wormholes.  In the limit
of $a=0$ when the wormhole disappears, the geodesic is complete at
$r=0$, and indeed the transmission rate vanishes.

\section{Wormholes in five-dimensional supergravities}

   In this section, we consider wormhole solutions in five-dimensional
$U(1)^3$ supergravity. This theory, sometimes called the STU model,
can be thought of as a truncation of maximal supergravity in five
dimensions, in which two vector multiplets are coupled to the minimal
supergravity.  It is well known that the theory admits magnetic black
string solutions.  Our Ricci-flat wormhole can be viewed as a neutral
tachyonic string.  It is natural to charge it as in the case of black
strings.  We find that the wormhole solution with the magnetic charges
is given by\footnote{We follow the convention of \cite{football} with
the cosmological constant turned off.}
\bea
ds^2 &=& -(H_1H_2H_3)^{-1/3}\Big(\fft{r^2-2as\, r -
a^2}{r^2 + a^2} dt^2 +\fft{4ac\,r}{r^2+a^2} dt\,dz -
\fft{r^2+2a s\, r-a^2}{r^2+a^2} dz^2\Big)\nn\\
&&+(H_1H_2H_3)^{2/3} \Big(dr^2 + (r^2 + a^2) d\Omega_2^2
\Big)\,,\nn\\
X_i&=&H_i (H_1H_2H_3)^{-1/3}\,,\qquad
F_\2^i=q_i\Omega_\2\,,\qquad
H_i=\alpha_i - \fft{q_i}{a} \arctan(\fft{r}{a})\,.
\eea
The solution is singular for any $H_i=0$, which can be avoided
provided that
\be
\alpha_i \ge \fft{\pi\, |q_i|}{2a}\,.\label{unequal}
\ee
Since the $H_i$'s approach constant when $r\rightarrow \pm \infty$,
the solution again describes wormholes that connect two asymptotic
flat spacetimes.  The essential features of the Ricci-flat wormhole
are retained, since the $H_i$'s are finite but non-vanishing.  In the
limit of $a=0$, the solution becomes the supersymmetric black string
whose near-horizon geometry is AdS$_3\times S^2$.  Indeed, it was
shown in \cite{ght} that dilatonless magnetic BPS $p$-branes have no
curvature singularity, and can be viewed as a wormhole, but with the
inner and outer horizon worlds identified.  In the BPS limit described
in section 2, the solution becomes a magnetically-charged string with
a pp-wave propagating along the string, with the near horizon geometry
as a direct product of the extremal BTZ black hole (locally AdS$_3$)
and $S^2$.

A particular interesting case arises when the equality in
(\ref{unequal}) is reached for all $i$.  In this case, in the
asymptotic region $r\rightarrow \infty$, we have $H_i\sim 1/r$.  It
follows that the metric is asymptotically AdS$_3\times S^2$.  On the
asymptotic region $r\rightarrow -\infty$, the $H_i$'s are constants,
and hence the metric is flat.  Thus the solution describes a wormhole
that connects a AdS$_3\times S^2$ at one end to the flat Minkoskian
spacetime in the other end.

Using the solution generating procedure that was developed in
\cite{lmp}, we can also easily charge the solution with electric
charge so that it becomes a solution of the $D=5$ minimal
supergravity.  Starting from the neutral wormhole (\ref{d5met1}), the
resulting solution is given by
\bea
ds^2&=& -\fft{f}{H^2} \Big( dt + 2ac(\fft{\td c^3 r dz}{r^2-a^2-2a r s} +
\td s^3 \cos\theta\,d\phi)\Big)^2\nn\\
&&+H \Big(dr^2 + (r^2 + a^2) (d\theta^2 + \sin^2\theta\,d\phi^2) +
\fft{1}{f} dz^2\Big)\,,\nn\\
A&=&\fft{2\sqrt3\,a \td c \td s}{H} \Big(
\fft{a + r s}{r^2 + a^2} dt-
\fft{r c \td c}{r^2 + a^2} dz-
c\td s\cos\theta\,d\phi\Big)\,,\nn\\
H&=& 1 + \fft{2a (a + r s) \td s^2}{r^2 + a^2}\,,\qquad
f=\fft{r^2 - a^2 - 2 a s r}{r^2 + a^2}\,.
\eea
Here $\td c$ and $\td s$ are the boost parameters satisfying $\td
c^2-\td s^2=1$.  The metric has a curvature power-law singularity when
$H=0$, but this can be avoided if the boosted parameter is less than a
critical value, namely
\be
\td s^2 <\fft{1 + c}{s^2}\,.
\ee
With this condition, $H$ is non-vanishing, and the solution describes
a charged wormhole.  The mass, charge and momentum are
given by
\be
M_\pm=\pm\ft34 a s (1 + 2 \td s^2)\,,\qquad
Q_e^\pm=\pm \ft12\sqrt3\, a s\, \td c \td s\,,\qquad
P_\pm=\pm\ft34 a c\, \td c^3\,.
\ee
In addition, there is a magnetic charge, given by
\be
Q_m=\fft{1}{8\pi}\int_{r\rightarrow \infty} F_2=\sqrt3\, ac \td c\,\td s^2\,.
\ee
The existence of this magnetic charge is responsible for that the metric
is not asymptotically flat, but has instead two G\"odel-like universes.
In the appendix, we present the general wormhole solution in $U(1)^3$
theory that carries both electric and magnetic charges.

\section{Conclusions}

In this paper we construct stationary Ricci-flat inter-universe
Lorentzian wormholes in all $D\ge 5$ dimensions.  We focus our
analysis on $D=5$.  The metric smoothly connects two asymptotic flat
spacetimes.  The solution can be viewed as supergravity dual of a
string tachyon state whose linear momentum is larger than its tension.
In the BPS limit, the solution becomes a supersymmetric pp-wave and
the two worlds are then disconnected by the pp-wave singularity.

The wormholes are not traversable for the timelike and null geodesics.
However, we show that there exist accelerated timelike trajectories
that traverse from one asymptotic region to the other.  The minimum
value of the maximum acceleration of various trajectories can be
arbitrarily small.  For the accelerated timelike trajectories that we
obtained, the total impulse required for cross over the wormhole is
finite and independent on the size of the wormhole, and its minimum
value is $\pi$.

     We further study the wave equation of the minimally-coupled
massive scalars.  We demonstrate that their quantum tunnelling
between the two worlds must take place.

We also obtain charged wormhole solutions in five-dimensional
supergravities.  Although the metric becomes more complicated, the
essential feature of the Ricci-flat wormholes is retained.  With
appropriate choice of parameters, we find that there exist wormholes
that connect AdS$_3\times S^2$ in one asymptotic region to flat
Minkoskian spacetime in the other.  It is of interest to study the
implication of such a geometry on the AdS/CFT correspondence.

Our results indicate that there exist a large class of smooth
Lorentzian wormhole solutions in higher dimensions that are supported
by less exotic energy-momentum tensors.  If we reduce our wormhole
solutions, with or without charges, on $t$ and $z$ directions, we
obtain a Euclidean wormhole in three dimensions supported by a scalar
coset.  General Euclidean wormholes in Euclidean supergravities
obtained from Kaluza-Klein reductions on tori that include the time
direction \cite{euclid} have been recently studied in \cite{bd,bcptv}.
These solutions typically have singularities associated with the
scalar coset, unless some liberal analytical continuation is performed
which may be inconsistent with supersymmetry and U-dualities \cite{bd}.
However, our results demonstrate that some of these lower-dimensional
singular solutions can become smooth Lorentzian wormholes when they
are lifted up to higher dimensions.

It is of great interest to classify these Lorentzian wormholes in
supergravities in higher dimensions and investigate their properties
such as the stability, geodesic traversability and the r\^ole they
play in string theory.

\section*{Acknowledgement}

We are grateful to Aaron Bergman, Gary Gibbons, Jianxin Lu, Chris
Pope, Don Page, Yi Pang and Yi Wang for useful discussions, and to Yi
Pang for reading the manuscript and correcting typos. Research is
supported in part by DOE grant DE-FG03-95ER40917.

\newpage

\appendix

\centerline{\Large\bf Appendix}

\section{Wormholes with both electric and magnetic charges}

Here we present the general wormhole solution in five-dimensional
$U(1)^3$ supergravity that carries both electric and magnetic charges.
It is given by
\bea%
ds^2&=&(h_1h_2h_3)^{1/3}(H_1H_2H_3)^{2/3}\left[dr^2+(a^2+r^2)d\Omega_2^2
  +\frac{dz^2}{H_1H_2H_3f}\right]\nonumber\\ &&-
\frac{f(dt+\omega)^2}{(h_1h_2h_3)^{2/3}(H_1H_2H_3)^{1/3}}~;\nonumber\\
\omega&=&c_1c_2c_3\left\{\left[f_z-\frac{t_2t_3}{H_1}-\frac{t_1t_3}{H_2}
  -\frac{t_1t_2}{H_3}\right]dz+(2act_1t_2t_3-t_1q_1-t_2q_2-t_3q_3)\cos\theta
d\phi\right\}~;\nonumber\\ A_i&=&
\frac{c_is_i\left[1-H_i^2f_t(r)\right]dt}{h_i}+\frac{c_1c_2c_3
  \left[t_jH_j+t_kH_k
    +(t_1t_2t_3-t_iH_if_z)f\right]}{h_iH_jH_k}dz\nonumber\\
&&+\frac{c_1c_2c_3\left[q_i-2act_jt_k+t_i(t_jq_j+t_kq_k)H_i^2f_t
    \right]}{h_i}\cos\theta d\phi~~~;~~~j\neq k\neq
i\nonumber\\ X_i&=&\frac{\left(h_1h_2h_3\right)^{1/3}H_i}{
\left(H_1H_2H_3\right)^{1/3}h_i}
~~~;~~~ H_i=a_i+\frac{q_i}{a}\tan^{-1}\left(\frac{r}{a}\right)~~~;~~~
h_i=c_i^2-s_i^2H_i^2f_t~;\nonumber\\ f_t&=&\frac{f}{H_1H_2H_3}~~~;~~~
f_z=\frac{2acr}{r^2-a^2-2ars} ~~~;~~~
f=\frac{r^2-a^2-2ars}{a^2+r^2}~~~;~~~t_i\equiv\frac{s_i}{c_i}~.
\eea%

\end{document}